\documentclass[10pt]{article}
\usepackage{multicol}
\usepackage{graphicx}
\usepackage{amsmath}
\usepackage[a4paper]{geometry}
\usepackage{CJK}

\begin{document}

\begin{center}
{\large\bf Energy dependent chemical potentials of light particles
and quarks\\ from yield ratios of antiparticles to particles in
high energy collisions}

\vskip0.75cm

Hai-Ling Lao$^1$, Ya-Qin Gao$^2$, Fu-Hu
Liu$^{1,}${\footnote{E-mail: fuhuliu@163.com; fuhuliu@sxu.edu.cn}}

\vskip0.2cm

{\small\it $^1$Institute of Theoretical Physics \& State Key
Laboratory of Quantum Optics and Quantum Optics Devices,

Shanxi University, Taiyuan, Shanxi 030006, China

$^2$Department of Physics, Taiyuan University of Science and
Technology, Taiyuan, Shanxi 030024, China}

\end{center}

\vskip0.5cm

{\bf Abstract:} We collect the yields of charged pions ($\pi^-$
and $\pi^+$), charged kaons ($K^-$ and $K^+$), anti-protons ($\bar
p$), and protons ($p$) produced in mid-rapidity interval (in most
cases) in central gold-gold (Au-Au), central lead-lead (Pb-Pb),
and inelastic or non-single-diffractive proton-proton ($pp$)
collisions at different collision energies. The chemical
potentials of light particles and quarks are extracted from the
yield ratios, $\pi^-/\pi^+$, $K^-/K^+$, and $\bar p/p$, of
antiparticles to particles over an energy range from a few GeV to
above 10 TeV. At a few GeV ($\sim4$ GeV), the chemical potentials
show, and the yield ratios do not show, different trends comparing
with those at other energies, though the limiting values of the
chemical potentials and the yield ratios at very high energy are 0
and 1 respectively.
\\



\vskip1.0cm

\begin{multicols}{2}

{\section{Introduction}}

Chemical potential ($\mu_{baryon}$) of baryon in high energy
collisions is a very interesting and important topic studied by
researchers in the fields of particle and nuclear physics.
Combining with temperature ($T_{ch}$) at chemical freeze-out, one
can study the quantum chromodynamics (QCD) phase diagram in the
plane of $T_{ch}$ against $\mu_{baryon}$ for the phase transition
from hadronic matter to quark-gluon plasma (QGP) [1--4]. It is
expected that this phase transition possibly happens over a
center-of-mass energy per nucleon pair, $\sqrt{s_{NN}}$, range
from a few GeV to dozens of GeV. The purpose of the beam energy
scan (BES) performed at the Super Proton Synchrotron (SPS) and the
Relativistic Heavy Ion Collider (RHIC) is to search for the
critical energy at which the phase transition from hadronic matter
to QGP had happened in all probability [5--8]. The BES energies at
the SPS reach or close to the Alternating Gradient Synchrotron
(AGS) energy.

Combining with the AGS, SPS (at its BES), and RHIC (at its BES),
one can study the QCD phase diagram over an energy range from a
few GeV to 200 GeV [1--4]. In particular, the Large Hadron
Collider (LHC) has extended the energy range to a few TeV and even
above 10 TeV [9--12]. It is convenient for researchers to study
the QCD phase transition further. At the same time, the excitation
functions of $T_{ch}$ and $\mu_{baryon}$ (the energy dependent
$T_{ch}$ and $\mu_{baryon}$) can be studied in the mentioned
energy range. Generally, the values of $T_{ch}$ and $\mu_{baryon}$
in given collisions can be obtained from the yield ratios of
antiparticles to particles in a given rapidity interval and
transverse momentum range. Although the chemical potentials of
other particles such as mesons can also be obtained from the yield
ratios of antiparticles to particles, few chemical potentials of
mesons have been studied in literature. This enlightens the
present work.

We are interested in the chemical potentials of different types of
particles in high energy collisions, which can be obtained from
the yield ratios of antiparticles to particles in a particular
form. We are also interested in the chemical potentials of
different flavors of quarks, which can also be obtained from the
same yield ratios of antiparticles to particles. It is expected
that chemical potentials of particles (or quarks) change with the
increase of collision energy. The excitation function (the
dependence on collision energy) of chemical potentials are
particularly interesting and worthy of study. From the chemical
potentials of particles (or quarks), we can evaluate the relative
densities of final particles (or produced quarks) at different
energies. These relative densities are useful in the understanding
of interacting mechanism. Because of the data of yield ratios
being very limited, we can only obtain the chemical potentials of
some particles and quarks conveniently.

In this paper, we collect the yields of charged pions ($\pi^-$ and
$\pi^+$), charged kaons ($K^-$ and $K^+$), anti-protons ($\bar
p$), and protons ($p$) produced in mid-rapidity interval (in most
cases) in central gold-gold (Au-Au) [3, 13--23], central lead-lead
(Pb-Pb) [24--29], and inelastic (INEL) or non-single-diffractive
(NSD) proton-proton ($pp$) collisions [3, 30--34] over a
center-of-mass energy per nucleon pair, $\sqrt{s_{NN}}$, range
from a few GeV to above 10 TeV. The chemical potentials of light
particles and quarks are extracted from the yield ratios,
$\pi^-/\pi^+$, $K^-/K^+$, and $\bar p/p$, of antiparticles to
particles, where the symbol of a given particle is used for its
yield for the purpose of simplicity. The energy dependent chemical
potentials of light particles and quarks are obtained due to the
yield ratios.
\\

{\section{The method and formalism}}

To extract the chemical potentials of light particles and quarks,
we need to know the yield ratios of antiparticles to particles.
Although we can obtain the yield ratios from the normalization
constants of transverse momentum spectra for different particles,
the quantity of work is huge if we analyze the spectra over a wide
energy range. A direct and convenient method is to collect the
values of yield ratios from the experiments performed at the AGS,
SPS at its BES, RHIC at its BES, and LHC by productive
international collaborations, though some yield ratios are not
available.

Because of the same formula on the relation between the yield
ratio and chemical potential being used in our previous work [35]
and the present work, some repetitions are ineluctable to give a
whole representation of the present work. According to the
statistical arguments based on the chemical and thermal
equilibrium, one has the relation between antiproton to proton
yield ratio to be [20, 36, 37]
\begin{equation}
\frac{\bar{p}}{p} =\exp\left( -\frac{2\mu_{p}}{T_{ch}}\right)
\approx \exp\left(-\frac{2\mu_{baryon}}{T_{ch}}\right)
\end{equation}
which is within the thermal and statistical model [36, 37], where
$\mu_p$ is the chemical potential of proton and
\begin{equation}
T_{ch}=T_{\lim}\frac{1}{1+\exp\left[ 2.60-\ln\left( \sqrt{s_{NN}}
\right)/0.45\right]}
\end{equation}
is empirically obtained in the framework of a statistical thermal
model of non-interacting gas particles with the assumption of
Boltzmann-Gibbs statistics [1, 2, 38, 39], where $\sqrt{s_{NN}}$
is in the units of GeV and the ``limiting" temperature
$T_{\lim}=0.164$ GeV.

We would like to point out that Eq. (1) obtained in the
statistical thermal model is due to the Boltzmann approximation
and the relation to isospin effect. Eq. (2) is due to the
Boltzmann approximation in the employ of grand-canonical ensemble,
though the employs of canonical ensemble and mix-strangeness
canonical ensemble being also considerable. As descriptions on
chemical freeze-out, Eqs. (1) and (2) do not include particles
with high transverse momenta ($>$2--3 GeV/$c$) which are produced
in hard scattering process at initial stage and leave the
interacting system before chemical freeze-out. The particles taken
part in chemical freeze-out should have low transverse momenta
($<$2--3 GeV/$c$) and obey the Boltzmann-Gibbs statistics.

According to Eq. (1), the yield ratios of antiparticles to
particles for other hadrons with together (anti)protons can be
written as
\begin{align}
k_j \equiv \frac{j^-}{j^+}
=\exp\left(-\frac{2\mu_j}{T_{ch}}\right),
\end{align}
where $k_{j}$ denotes the yield ratio of antiparticles to
particles of the kind $j$, and $j=\pi$, $K$, $p$, $D$, and $B$
listed orderly due to their masses. The symbol $\mu_{j}$
represents the chemical potential of particle $j$. To obtain
chemical potentials of quarks, the above five hadrons and their
antiparticles are enough. Because of the lifetimes of particles
contained top quark being very short to measure, we shall not
discuss the top quark related particles, top quark itself, and
their chemical potentials.

Let $\mu_{q}$ denote the chemical potential for quark flavor,
where $q=u$, $d$, $s$, $c$, and $b$ represent the up, down,
strange, charm, and bottom quarks, respectively. The values of
$\mu_{q}$ are then expected from these relations. According to
refs. [40, 41], based on the same chemical freeze-out temperature,
the yield ratios in terms of quark chemical potentials are
\begin{align}
k_{\pi} & =\exp\left[ -\frac{\left( \mu_{u}-\mu_{d}\right)}
{T_{ch}}\right] \bigg/\exp\left[  \frac{\left(
\mu_{u}-\mu_{d}\right)} {T_{ch}}\right] \nonumber\\ & =\exp\left[
-\frac{2\left(\mu_{u}-\mu_{d}\right)}
{T_{ch}}\right],\nonumber\\
k_{K} & =\exp\left[ -\frac{\left( \mu_{u}-\mu_{s}\right)}
{T_{ch}}\right] \bigg/\exp\left[ \frac{\left(
\mu_{u}-\mu_{s}\right)} {T_{ch}}\right] \nonumber\\ & =\exp\left[
-\frac{2\left( \mu_{u}-\mu_{s}\right)}
{T_{ch}}\right],\nonumber\\
k_{p} & =\exp\left[ -\frac{\left( 2\mu_{u}+\mu_{d}\right)}
{T_{ch}}\right] \bigg/\exp\left[ \frac{\left(
2\mu_{u}+\mu_{d}\right)} {T_{ch}}\right] \nonumber\\ & =\exp\left[
-\frac{2\left(2\mu_{u}+\mu_{d}\right)}
{T_{ch}}\right],\nonumber\\
k_{D} & =\exp\left[ -\frac{\left( \mu_{c}-\mu_{d}\right)}
{T_{ch}}\right] \bigg/\exp\left[ \frac{\left(
\mu_{c}-\mu_{d}\right)} {T_{ch}}\right] \nonumber\\ & =\exp\left[
-\frac{2\left( \mu_{c}-\mu_{d}\right)}
{T_{ch}}\right],\nonumber\\
k_{B} & =\exp\left[ -\frac{\left( \mu_{u}-\mu_{b}\right)}
{T_{ch}}\right] \bigg/\exp\left[ \frac{\left(
\mu_{u}-\mu_{b}\right)} {T_{ch}}\right] \nonumber\\ & =\exp\left[
-\frac{2\left( \mu_{u}-\mu_{b}\right)} {T_{ch}}\right].
\end{align}

According to Eqs. (3) and (4), the chemical potentials of
particles and quarks can be obtained, respectively, in terms of
yield ratios of antiparticles to particles. The chemical potential
of particle $j$ is simply given by
\begin{align}
\mu_j =-\frac{1}{2}T_{ch}\cdot\ln\left( k_j\right).
\end{align}
The chemical potentials of quarks $q$ are complicatedly a little.
We have
\begin{align}
\mu_{u} & =-\frac{1}{6}T_{ch}\cdot \ln\left( k_{\pi}\cdot
k_{p}\right),\nonumber\\
\mu_{d} & =-\frac{1}{6}T_{ch}\cdot \ln\left( k_{\pi}^{-2}\cdot
k_{p}\right),\nonumber\\
\mu_{s} & =-\frac{1}{6}T_{ch}\cdot \ln\left( k_{\pi}\cdot
k_{K}^{-3}\cdot k_{p}\right),\nonumber\\
\mu_{c} & =-\frac{1}{6}T_{ch}\cdot \ln\left( k_{\pi}^{-2}\cdot
k_{p}\cdot k_{D}^{3}\right),\nonumber\\
\mu_{b} & =-\frac{1}{6}T_{ch}\cdot \ln\left( k_{\pi}\cdot
k_{p}\cdot k_{B}^{-3}\right).
\end{align}

Because of the limited data in extracting chemical potentials of
some quarks, only the energy dependent chemical potentials of
light particles such as $\pi$, $K$, and $p$, as well as light
quarks such as $u$, $d$, and $s$ in an energy range covered the
AGS, SPS (at its BES), RHIC (at its BES), and LHC are obtained in
the present work. That is to say that, in the present work, only
the excitation functions of $\mu_{\pi}$, $\mu_K$, $\mu_p$,
$\mu_u$, $\mu_d$, and $\mu_s$ are studied over an energy range
from a few GeV to above 10 TeV. For central Au-Au (Pb-Pb)
collisions and INEL or NSD $pp$ colliisons, the energy ranges are
not completely corresponding to each other.

It should be noted that other yield ratios such as
$\bar\Lambda/\Lambda$, $\bar\Sigma/\Sigma$, and
$\bar\Omega/\Omega$ do extract only chemical potentials of light
quarks due to the fact that their constituent quarks are only
light quarks, which are not needed in the present work in
particular. Although these yield ratios are available in
experiments at some energies, they will not be analyzed by us.
\\

\begin{figure*}[!htb]
\begin{center}
\includegraphics[width=15.0cm]{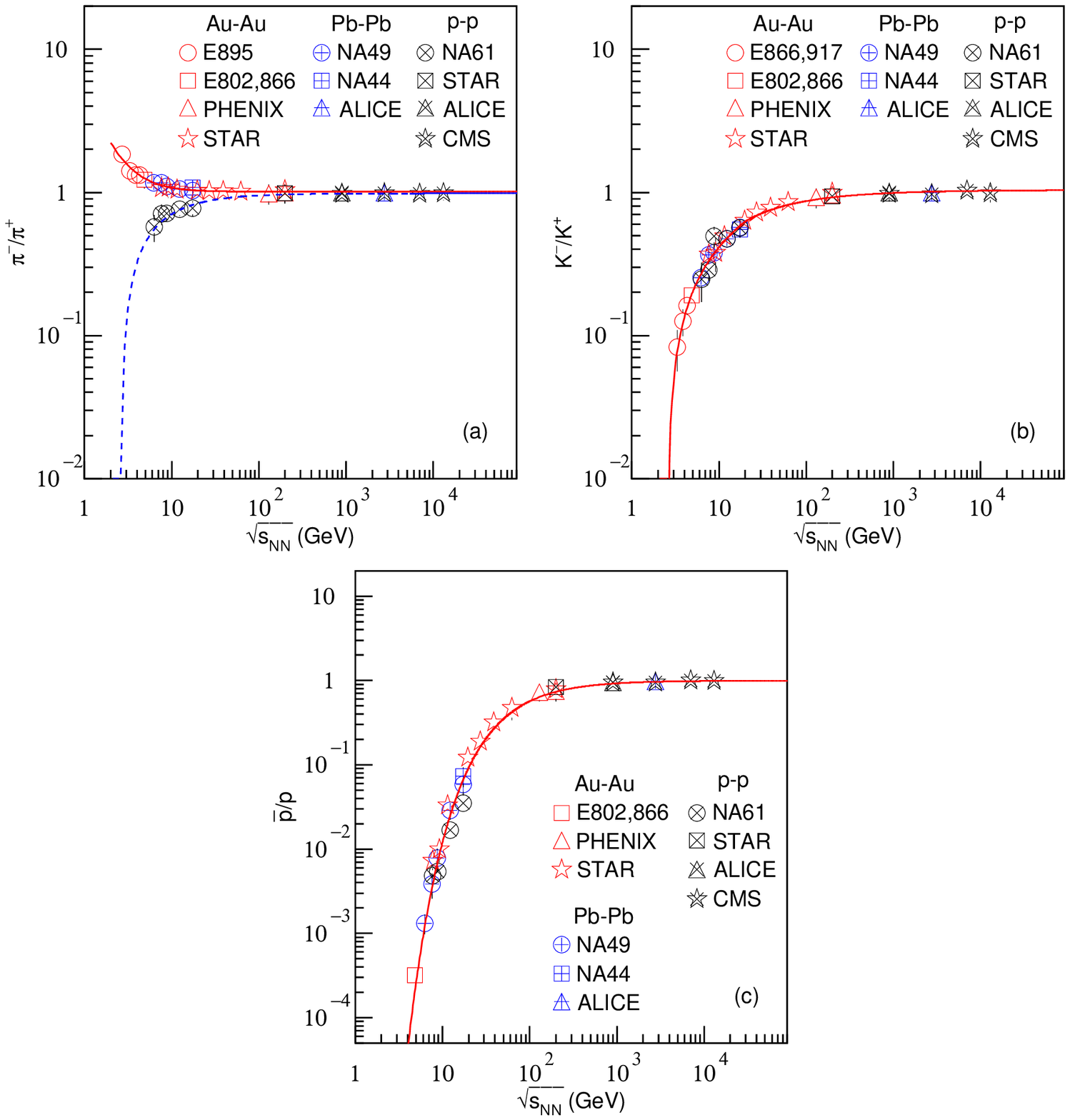}
\end{center}
Fig. 1: Yield ratios, (a) $\pi^-/\pi^+$, (b) $K^-/K^+$, and (c)
$\bar p/p$, of antiparticles to particles produced in mid-rapidity
interval (in most cases) in central Au-Au (Pb-Pb) and INEL or NSD
$pp$ collisions at high energies. The symbols denote the data
measured in different collisions by different collaborations
marked in the panels, where the idiographic references are indexed
in the text. In particular, the NA61/SHINE data appear in the
forward rapidity region (in the center-of-mass system, though the
experiment can provide results with $4\pi$ geometry. The solid and
dashed curves in Fig. 1(a) are the results fitted by us for the
$\sqrt{s_{NN}}$ dependent $\pi^-/\pi^+$ in central Au-Au (Pb-Pb)
and INEL or NSD $pp$ collisions respectively. The solid curves in
Figs. 1(b) and 1(c) are the results fitted by us for the
$\sqrt{s_{NN}}$ dependent $K^-/K^+$ and $\bar p/p$ respectively,
for the combining central Au-Au (Pb-Pb) and INEL or NSD $pp$
collisions.
\end{figure*}

{\section{Results and discussion}}

The yield ratios, $\pi^-/\pi^+$, $K^-/K^+$, and $\bar p/p$, of
antiparticles to particles produced in mid-rapidity interval (in
most cases) in central Au-Au, central Pb-Pb, and INEL or NSD $pp$
collisions at the AGS, SPS, RHIC, and LHC are shown in Figs. 1(a),
1(b), and 1(c), respectively. The circles, squares, triangles, and
starts denote the data measured in Au-Au collisions in
mid-rapidity interval from $|y|<0.05$ to $|y|<0.4$ and centrality
0--5\% by the E895, E866, and E917 Collaborations [13--15] at the
AGS, in mid-rapidity interval $|y|<0.4$ and centrality 0--10\% by
the E802 and E866 Collaboration [16, 17] at the AGS, in
mid-pseudorapidity interval $|\eta|<0.35$ and centrality 0--5\% by
the PHENIX Collaboration [18--20], and in mid-rapidity interval
from $|y|<0.1$ to $|y|<0.5$ and centrality from 0--5\% to 0--10\%
by the STAR Collaborations [3, 21--23] at the RHIC, respectively.
The circles, squares, and triangles with aclinal crosses denote
the data measured in Pb-Pb collisions in mid-rapidity interval
from $0<y<0.2$ or $|y|<0.1$ to $|y|<0.6$ and centrality from
0--5\% to 0--7.2\% by the NA49 Collaboration [24--27] at the SPS,
in mid-rapidity interval from $|y|<0.5$ to $|y|<0.85$ and
centrality 0--3.7\% by the NA44 Collaboration [28] at the SPS, and
in mid-rapidity interval $|y|<0.5$ and centrality 0--5\% by the
ALICE Collaboration [29] at the LHC, respectively. The circles,
squares, triangles, and stars with diagonal crosses denote the
data measured in the forward rapidity region (in the
center-of-mass system) in INEL $pp$ collisions by the NA61/SHINE
Collaboration [30] at the SPS, in mid-rapidity interval $|y|<0.1$
in NSD $pp$ collisions by the STAR Collaboration [3, 31] at the
RHIC, in mid-rapidity interval $|y|<0.5$ in INEL $pp$ collisions
by the ALICE Collaboration [32] at the LHC, and in mid-rapidity
interval $|y|<1$ in INEL $pp$ collisions by the CMS Collaboration
[33, 34] at the LHC, respectively. The solid and dashed curves in
Fig. 1(a) are the results fitted by us for the $\sqrt{s_{NN}}$
dependent $\pi^-/\pi^+$ in central Au-Au (Pb-Pb) and INEL or NSD
$pp$ collisions respectively. The solid curves in Figs. 1(b) and
1(c) are the results fitted by us for the $\sqrt{s_{NN}}$
dependent $K^-/K^+$ and $\bar p/p$ respectively, for the combining
central Au-Au (Pb-Pb) and INEL or NSD $pp$ collisions.

It should be noted that the $\pi^-/\pi^+$ ratios presented in Fig.
1(a) obtained by the NA61/SHINE Collaboration from $pp$ collisions
at around 10 GeV are so different to the others obtained from
Au-Au (Pb-Pb) collisions at the same energy. This is due to the
resonance decay existed mainly in nucleus-nucleus collisions over
an energy range from a few GeV to dozens of GeV [42]. There are
secondary cascade collisions between produced particles and
subsequent nucleons in nucleus-nucleus collisions, which can
produce resonances which then decay and affect mainly
$\pi^-/\pi^+$ ratios. For $K^-/K^+$ and $\bar p/p$ ratios
presented in Figs. 1(b) and 1(c) respectively, the effect of
resonance decay in nucleus-nucleus collisions is not obvious. At
very high energy (above 100 GeV), the contribution of resonances
can be neglected, which renders similar results in $pp$ and Au-Au
(Pb-Pb) collisions. In addition, more energies are deposited in
Au-Au (Pb-Pb) collisions than in $pp$ collisions, which also
results the difference in $\pi^-/\pi^+$ ratios. Meanwhile, the
deposited energies are not too large, which does not result in the
difference in $K^-/K^+$ ($\bar p/p$) ratios.

One can see from Fig. 1 that, with the increase of
$\sqrt{s_{NN}}$, $\pi^-/\pi^+$ decreases obviously in central
Au-Au (Pb-Pb) collisions and it increases obviously in INEL or NSD
$pp$ collisions, and $K^-/K^+$ and $\bar p/p$ increase obviously
in both central Au-Au (Pb-Pb) and INEL or NSD $pp$ collisions. The
limiting values of the three yield ratios is 1 at very high
energy. The solid and dashed curves in Fig. 1(a) can be
empirically described by
\begin{align}
\frac{\pi^-}{\pi^+}=&(4.212\pm0.682)\cdot(\sqrt{s_{NN}})^{-(1.799\pm0.152)} \nonumber \\
&+(1.012\pm0.019)
\end{align}
and
\begin{align}
\frac{\pi^-}{\pi^+}=&-(2.453\pm0.292)\cdot(\sqrt{s_{NN}})^{-(0.943\pm0.057)} \nonumber \\
&+(0.984\pm0.009)
\end{align}
respectively, with $\chi^2$/dof ($\chi^2$ per degree of freedom)
to be 0.162 and 1.559 respectively. The solid curves in Figs. 1(b)
and 1(c) can be empirically described by
\begin{align}
\frac{K^-}{K^+}=&\big[-(0.291\pm0.028)+(0.306\pm0.010)\cdot\ln(\sqrt{s_{NN}})\big] \nonumber \\
&\cdot \theta(20-\sqrt{s_{NN}}) \nonumber \\
&+\big[-(2.172\pm0.146)\cdot(\sqrt{s_{NN}})^{-(0.554\pm0.018)} \nonumber \\
&\hskip4mm +(1.039\pm0.016)\big] \nonumber \\
&\cdot\theta(\sqrt{s_{NN}}-20)
\end{align}
and
\begin{align}
\frac{\bar p}{p}=&\exp\big[-(34.803\pm3.685)\cdot(\sqrt{s_{NN}})^{-(0.896\pm0.041)}  \nonumber \\
&-(0.008\pm0.004)\big]
\end{align}
respectively, with $\chi^2$/dof to be 2.735 and 7.715
respectively. According to these functions, by using Eqs. (5) and
(6), the chemical potentials of light particles and quarks can be
obtained.

\begin{figure*}[!htb]
\begin{center}
\includegraphics[width=15.0cm]{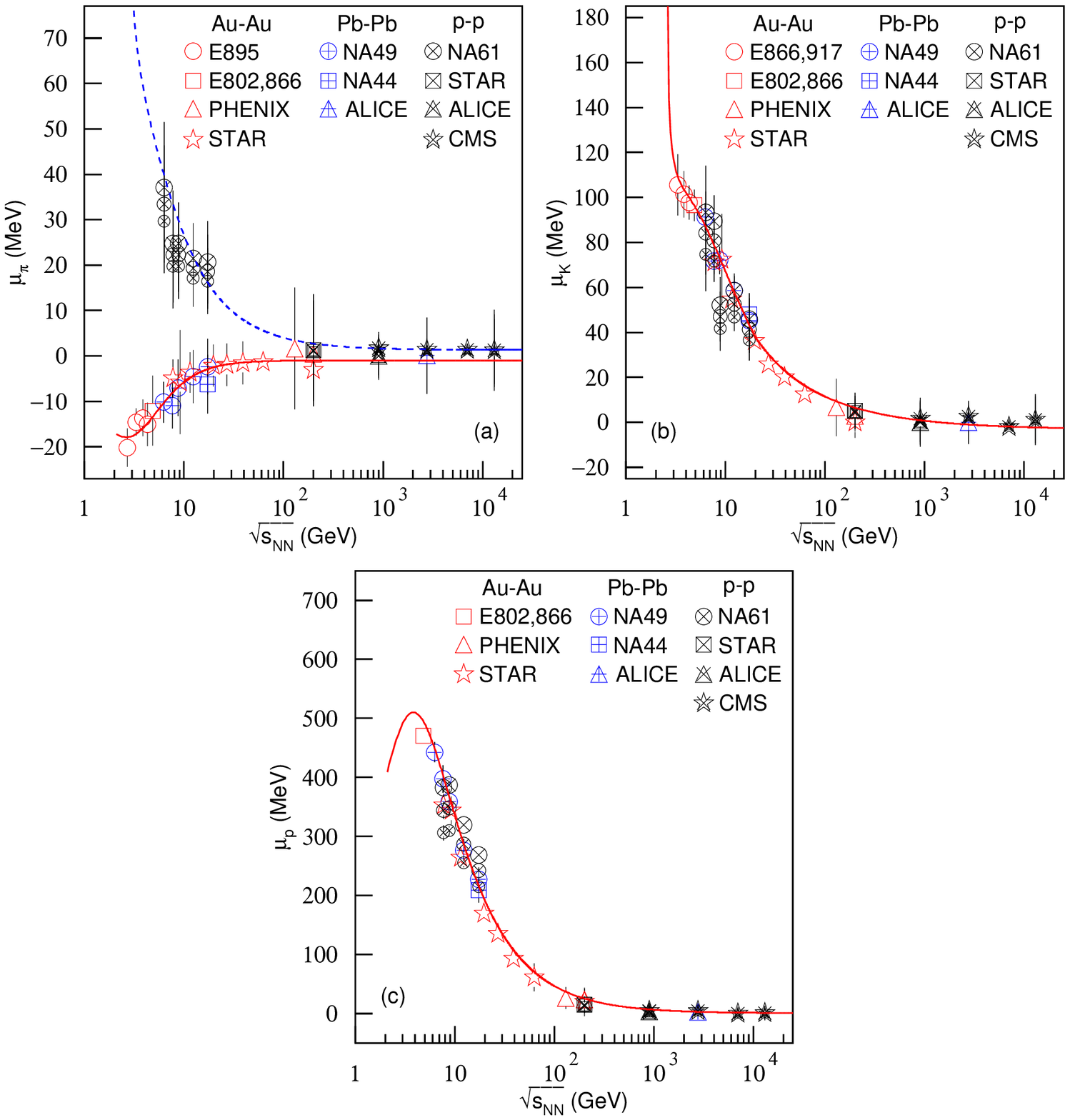}
\end{center}
Fig. 2: Chemical potentials, (a) $\mu_{\pi}$, (b) $\mu_K$, and (c)
$\mu_p$, of (a) $\pi$, (b) $K$, and (c) $p$ produced in
mid-rapidity interval (in most cases) in central Au-Au (Pb-Pb) and
INEL or NSD $pp$ collisions at high energies. The symbols denote
the derivative data obtained from Fig. 1 according to Eq. (5). In
particular, the NA61/SHINE data appear in the forward rapidity
region (in the center-of-mass system), though the experiment can
provide results with $4\pi$ geometry. The normal, medium, and
small symbols with diagonal crosses denote the derivative data in
INEL or NSD $pp$ collisions obtained by $T_{ch}$, $0.9T_{ch}$, and
$0.8T_{ch}$ in Eq. (5), respectively. The curves surrounded the
symbols are the derivative results obtained from the curves in
Fig. 1 according to Eq. (5).
\end{figure*}

\begin{figure*}[!htb]
\begin{center}
\includegraphics[width=15.0cm]{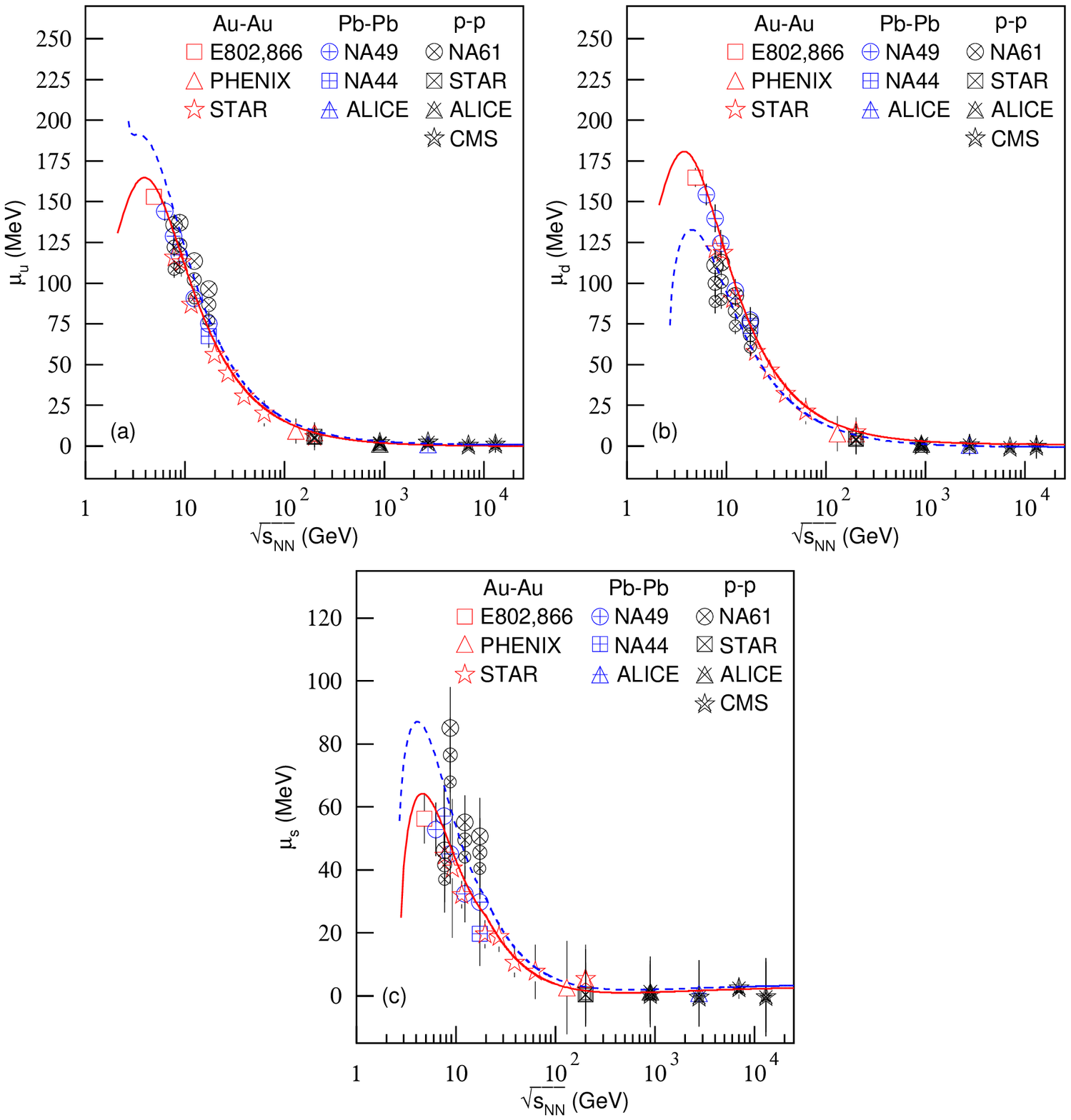}
\end{center}
Fig. 3: Chemical potentials, (a) $\mu_u$, (b) $\mu_d$, and (c)
$\mu_s$, of (a) $u$, (b) $d$, and (c) $s$ quarks derived from
mid-rapidity interval (in most cases) in central Au-Au (Pb-Pb) and
INEL or NSD $pp$ collisions at high energies. The symbols denote
the derivative data obtained from Fig. 1 according to Eq. (6). In
particular, the NA61/SHINE data appear in the forward rapidity
region (in the center-of-mass system), though the experiment can
provide results with $4\pi$ geometry. The normal, medium, and
small symbols with diagonal crosses denote the derivative data in
INEL or NSD $pp$ collisions obtained by $T_{ch}$, $0.9T_{ch}$, and
$0.8T_{ch}$ in Eq. (6), respectively. The curves surrounded the
symbols are the derivative results obtained from the curves in
Fig. 1 according to Eq. (6).
\end{figure*}

Figures 2(a), 2(b), and 2(c) present respectively the chemical
potentials, $\mu_{\pi}$, $\mu_K$, and $\mu_p$, of $\pi$, $K$, and
$p$ produced in mid-rapidity interval (in most cases) in central
Au-Au (Pb-Pb) and INEL or NSD $pp$ collisions at high energies.
The symbols denote the derivative data obtained from Fig. 1
according to Eq. (5), where different symbols correspond to
different collaborations marked in the panels which are the same
as Fig. 1. For the purpose of comparison, the normal, medium, and
small symbols with diagonal crosses denote the derivative data in
INEL or NSD $pp$ collisions obtained by $T_{ch}$, $0.9T_{ch}$, and
$0.8T_{ch}$ in Eq. (5), respectively, due to the fact that the
chemical freeze-out temperature in $pp$ collisions is not
available. The curves are the derivative results obtained from the
curves in Fig. 1 according to Eq. (5). The solid and dashed curves
in Fig. 2(a) are the derivative results for central Au-Au (Pb-Pb)
and INEL or NSD $pp$ collisions respectively. The solid curves in
Figs. 2(b) and 2(c) are the derivative results for the combining
central Au-Au (Pb-Pb) and INEL or NSD $pp$ collisions. One can see
that, with the increase of $\sqrt{s_{NN}}$ over a range from above
a few GeV to above 10 TeV, $\mu_{\pi}$ increases obviously in
central Au-Au (Pb-Pb) collisions and it decreases obviously in
INEL or NSD $pp$ collisions, and $\mu_K$ and $\mu_p$ decrease
obviously in both central Au-Au (Pb-Pb) and INEL or NSD $pp$
collisions. The limiting values of the three types of chemical
potentials are 0 at very high energy. As derivative results from
Fig. 1, the difference and similarity in $pp$ and Au-Au (Pb-Pb)
collisions are natural due to the application of Eq. (5).

Figure 3 is the same as Fig. 2, but Figs. 3(a), 3(b), and 3(c)
present respectively the chemical potentials, $\mu_u$, $\mu_d$,
and $\mu_s$, of $u$, $d$, and $s$ quarks, which are derived from
mid-rapidity interval (in most cases) in central Au-Au (Pb-Pb) and
INEL or NSD $pp$ collisions at high energies. The symbols denote
the derivative data obtained from Fig. 1 according to Eq. (6),
where different symbols correspond to different collaborations
marked in the panels which are the same as Figs. 1 and 2. For the
purpose of comparison, the normal, medium, and small symbols with
diagonal crosses denote the derivative data in INEL or NSD $pp$
collisions obtained by $T_{ch}$, $0.9T_{ch}$, and $0.8T_{ch}$ in
Eq. (6), respectively. The curves are the derivative results
obtained from the curves in Fig. 1 according to Eq. (6), where the
solid and dashed curves are for central Au-Au (Pb-Pb) and INEL or
NSD $pp$ collisions respectively. One can see that, with the
increase of $\sqrt{s_{NN}}$ over a range from above a few GeV to
above 10 TeV, $\mu_u$, $\mu_d$, and $\mu_s$ decrease obviously in
both central Au-Au (Pb-Pb) and INEL or NSD $pp$ collisions. The
limiting values of the three chemical potentials are 0 at very
high energy.

In Fig. 2, at a few GeV ($\sim4$ GeV), some curves show different
trends comparing with those at other energies, which are not
observed from the curves of the yield ratios in Fig. 1. At the
same time, the curves in Fig. 3 also show different trends at a
few GeV. Indeed, this energy is a special energy. In our opinion,
these special trends appear due to this energy being the initial
energy of limiting fragmentation of collision nuclei. This energy
is also the energy at which the phase transition from a
liquid-like state to a gas-like state in the collision system is
expected to happen initially, where the liquid-like state is a
state in which the mean-free-path of interacting particles is
relatively short, and the gas-like state is a state in which the
mean-free-path of interacting particles is relatively long. In
addition, the density of baryon number in nucleus-nucleus
collisions at this energy has a large value. Because of these
particular factors, the collisions at this energy present
different features from other energies. The matter formed at this
energy changes initially its state from the liquid-like nucleons
and mesons to the gas-like nucleons and mesons in whole stage of
collisions.

The above explanation on the initial energy of limiting
fragmentation of collision nuclei or the energy of the
phase-transition from the liquid-like state to the gas-like state
in the collision system is deservedly to discuss. In fact, the
present work contains rather standard analysis, performed since
many years [43--46] without too much novel results in the field.
The conclusion concerning physics presented in the present work is
possibly too far-reaching. Instead, the yield ratios discussed in
the present work can be also explained within the statistical
hadron resonance gas (HRG) model and the ultrarelativistic quantum
molecular dynamics (UrQMD) transport model [43--46]. At a few GeV,
the collision system stays at the state with the maximum density
and minimum radius, which results different trends of chemical
potentials.

From above a few GeV to dozens of GeV, the cases of $k_{\pi}>1$
and $\mu_{\pi}<0$ in central Au-Au (Pb-Pb) collisions are
different from other particles and in INEL or NSD $pp$ collisions.
These render the resonant production of pions in central Au-Au
(Pb-Pb) collisions, which does not contribute too much to other
particles or in INEL or NSD $pp$ collisions [42]. At the RHIC and
LHC, the trends of $k_{\pi}$ and $\mu_{\pi}$ in central Au-Au
(Pb-Pb) collisions are close to other particles or INEL or NSD
$pp$ collisions due to the insignificant contribution to pions and
other particles. From above a few GeV to above 10 TeV, the yield
ratios approach to 1 and the chemical potentials approach to 0.
These render that the mean-free-path of produced particles
(quarks) becomes large and the viscous effect becomes weakly at
the LHC. The interacting system changes completely from the
liquid-like state which is hadron-dominant to the gas-like state
which is quark-dominant at the early and medium stage in
collisions at very high energy, though the final stage is
hadron-dominant gas-like state.

Before giving conclusions, we discuss further the extraction
method of chemical potentials. Theoretically, chemical potentials
always correspond to some conserved charge. For example, in Ref.
[39], it is written how a hadron $i$ has a chemical potential. One
has $\mu_i = \mu_{baryon} B_i + \mu_S S_i + \mu_I I_i + \mu_C
C_i$, where $\mu$ with a lower foot mark corresponds to each
chemical potential, and $B_i$, $S_i$, $I_i$, and $C_i$ are the
particle's baryon number, strangeness, isospin, and charm,
respectively. Although not all of them are free parameters since
some of them are fixed by the conservation laws and some of them
are 0 for a special particle, it is still lesser-known to
determine $\mu_i$. In particular, to determine the chemical
potentials of quarks is more lesser-known.

To determine the chemical potentials, the present work tries to
use a convenient method. In the case of utilizing $T_{ch}$,
chemical potentials are obtained according to the yield ratios of
antiparticles to particles. In the extraction, the difference
between the chemical potentials of antiparticles and particles is
neglected, and the difference between the chemical potentials of
quark and its anti-quark is also neglected. Then, Eqs. (1), (3),
and (4) are acceptable. In addition, we have used a
single-$T_{ch}$ scenario for the chemical freeze-out, though a
two-$T_{ch}$ (or multi-$T_{ch}$) scenario is also possible. In
most cases, one intends to use the single-$T_{ch}$ scenario for
the chemical freeze-out.
\\

{\section{Conclusions}}

In summary, we have collected the yield ratios, $\pi^-/\pi^+$,
$K^-/K^+$, and $\bar p/p$, of antiparticles to particles produced
in mid-rapidity interval (in most cases) in central Au-Au (Pb-Pb)
and INEL or NSD $pp$ collisions over an energy range from a few
GeV to above 10 TeV. It is shown that, with the increase of
$\sqrt{s_{NN}}$, $\pi^-/\pi^+$ decreases obviously in central
Au-Au (Pb-Pb) collisions and it increases obviously in INEL or NSD
$pp$ collisions, and $K^-/K^+$ and $\bar p/p$ increase obviously
in both central Au-Au (Pb-Pb) and INEL or NSD $pp$ collisions. The
limiting values of the three yield ratios is 1 at very high
energy.

The chemical potentials of light particles and quarks are
extracted from the yield ratios. With the increase of
$\sqrt{s_{NN}}$ over a range from above a few GeV to above 10 TeV,
$\mu_{\pi}$ increases obviously in central Au-Au (Pb-Pb)
collisions and it decreases obviously in INEL or NSD $pp$
collisions, and $\mu_K$ and $\mu_p$ decrease obviously in both
central Au-Au (Pb-Pb) and INEL or NSD $pp$ collisions. Meanwhile,
$\mu_u$, $\mu_d$, and $\mu_s$ decrease obviously in both central
Au-Au (Pb-Pb) and INEL or NSD $pp$ collisions. The limiting values
of the chemical potentials of the three types of light particles
and the three flavors of light quarks are 0 at very high energy.

At a few GeV ($\sim4$ GeV), some curves of the chemical potentials
show different trends which are not observed from the curves of
the yield ratios. These special trends appear due to this energy
being possibly the initial energy of limiting fragmentation of
collision nuclei. This energy is also the energy of the phase
transition from the liquid-like state to the gas-like state in the
collision system. The interacting system changes completely from
the hadron-dominant liquid-like state to the quark-dominant
gas-like state at the early and medium stage in collisions at very
high energy.

The yield ratios discussed in the present work can also be
explained within the statistical hadron resonance gas model and
the ultrarelativistic quantum molecular dynamics transport model
[43--46]. At a few GeV, the collision system stays at the state
with the maximum density and minimum radius, which results
different trends of chemical potentials. The density of baryon
number in nucleus-nucleus collisions at a few GeV has a large
value. These particular factors render different features at this
energy.
\\

{\bf Data Availability}

The data used to support the findings of this study are quoted
from the mentioned references. As a phenomenological work, this
paper does not report new data.
\\

{\bf Conflicts of Interest}

The authors declare that there are no conflicts of interest
regarding the publication of this paper.
\\

{\bf Acknowledgments}

This work was supported by the National Natural Science Foundation
of China under Grant Nos. 11575103, 11747319, and 11847311, the
Scientific and Technological Innovation Programs of Higher
Education Institutions in Shanxi (STIP) under Grant No. 201802017,
the Shanxi Provincial Natural Science Foundation under Grant No.
201701D121005, the Fund for Shanxi ``1331 Project" Key Subjects
Construction, and the Doctoral Scientific Research Foundation of
Taiyuan University of Science and Technology under Grant No.
20152043.
\\

\end{multicols}
\end{document}